\documentclass[12pt]{article}
\usepackage{euscript,amsmath, amssymb, amsfonts}

\newcommand{\bee}{\begin{eqnarray}}
\newcommand{\eee}{\end{eqnarray}}
\newcommand{\nn}{\nonumber \\}
\newcommand{\R}{{\mathbb R}}
\newcommand{\oC}{{\mathbb C}}
\newcommand{\di}{{\rm d}}
\newcommand{\D}{\EuScript D}
\newcommand{\G}{{\mathbb G}}
\newcommand{\K}{{\mathbb K}}
\newcommand{\be}{\begin{equation}}
\newcommand{\ee}{\end{equation}}

\font\frtnfr=eufm10   scaled\magstep1 \font\twlfr=eufm10 \font\tenfr=eufm10
 
\newfam\frfam
\textfont\frfam=\frtnfr \scriptfont\frfam=\twlfr
\scriptscriptfont\frfam=\tenfr

\font\frtnopen=msbm10  scaled\magstep2 \font\twlopen=msbm10
\font\tenopen=msbm10  
\newfam\openfam
\textfont\openfam=\frtnopen \scriptfont\openfam=\twlopen
\scriptscriptfont\openfam=\tenopen
\def\open{\fam\openfam}

\font\frtnsf = cmss12 scaled\magstep1 \font\twlsf = cmss10 \font\tensf =
cmss9  
\newfam\Scfam
\textfont\Scfam = \frtnsf \scriptfont\Scfam = \twlsf \scriptscriptfont\Scfam
= \tensf

\makeatletter \@addtoreset{equation}{section}

\makeatother

\begin{document}

\sloppy \title
 {
Cohomologies of superalgebra
of pointwise superproduct
 }
\author
 {
 S.E.Konstein\thanks{E-mail: konstein@lpi.ru}\ \ and
 I.V.Tyutin\thanks{E-mail: tyutin@lpi.ru}
 \thanks{
               This work was supported
               by the RFBR (grant  No.~05-02-17217),
               and by the grant LSS-4401.2006.2.
 } \\
               {\sf \small I.E.Tamm Department of
               Theoretical Physics,} \\ {\sf \small P. N. Lebedev Physical
               Institute,} \\ {\sf \small 119991, Leninsky Prospect 53,
               Moscow, Russia.} }
\date {}

\maketitle

\begin{abstract}
{ \footnotesize We consider associative superalgebra realized on the smooth
Grassmann-valued functions with compact supports in $\R^n$. The lower Hochschild
cohomologies of this superalgebra
are found.
}
\end{abstract}


\section{Introduction}
\label{intr}

The hope to construct the quantum mechanics on nontrivial manifolds is
connected with geometrical or deformation quantization \cite{1} - \cite{Kon}.
The functions on the phase space are associated with the operators, and the
product and the commutator of the operators are described by associative
$*$--product and $*$--commutator of the functions.
These $*$-product and $*$-commutator
are the deformations of usual product and of usual Poisson bracket.

The gauge theories on the noncommutative
spaces, the so--called noncommutative gauge theories,  are formulated in
terms of the $*$--product (see \cite {Nek} and \cite {AS} and references
wherein).

The structure and the properties of the $*$--product on the usual (even)
manifolds are
investigated in details \cite{KarMas}, \cite{Fed3} and \cite{Kon}.
On the other hand, the $*$--product on supermanifolds is not
investigated  sufficiently.

It is easy to  extend formally the noncommutative product proposed
in \cite{Gro} to the supercase \cite{Ber}, however, there is
a problem of the uniqueness of
the $*$--product (the uniqueness of the "pointwise" product deformation).
In \cite{Tu}, the general form of the
associative $*$--product, treated as a deformation of the "pointwise"
product, on the Grassman algebra of a finite number of generators, is found
and uniqueness is proved.
It is interesting to solve analogous problem for functions of even and odd variables.

Each $*$--product generates $*$--commutators, but converse is not true (see
discussion in \cite{Leites}). It seems that non Moyal
deformations of Poisson superbracket found in \cite{TK} are not generated by
$*$--product.

The problem of finding $*$--products is connected with the problem
of calculating Hochschild cohomologies.
The Hochschild cohomologies of the algebra of smooth functions on ${\open R}^n$
are calculated in \cite{Zh1}.

Below we calculate lower Hochschild cohomologies (up to 2-nd)
of the algebra of smooth
Grassmann valued functions with compact supports on ${\open R}^n$.
Because we need these cohomologies for finding $*$--product, which is
even 2-form, we consider only even cocycles in this paper.

\section{Notation}
Let $\K$ be either $\R$ or $\oC$. We denote by $\EuScript D(\R^n)$ the space
of smooth $\K$-valued functions with compact support on $\R^n$. This space is
endowed with its standard topology: by definition, a sequence $\varphi_k\in
\EuScript D(\R^n)$ converges to $\varphi\in \EuScript D(\R^n)$ if the
supports of all $\varphi_k$ are contained in a fixed compact set, and
$\partial^\lambda\varphi_k$ converge uniformly to $\partial^\lambda\varphi$
for every multi-index $\lambda$. We set
$$
\mathbf D^{n_-}_{n_+}= \EuScript D(\R^{n_+})\otimes \G^{n_-},\quad \mathbf
D^{\prime n_-}_{n_+}= \EuScript D'(\R^{n_+})\otimes \G^{n_-},
$$
where $\G^{n_-}$ is the Grassmann algebra with $n_-$ generators and
$\EuScript D'(\R^{n_+})$ is the space of continuous linear functionals on
$\EuScript D(\R^{n_+})$. The generators of the Grassmann algebra (resp., the
coordinates of the space $\R^{n_+}$) are denoted by $\xi^\alpha$,
$\alpha=1,\ldots,n_-$ (resp., $x^i$, $i=1,\ldots, n_+$). We shall also use
collective variables $z^A$ which are equal to $x^A$ for $A=1,\ldots,n_+$ and
are equal to $\xi^{A-n_+}$ for $A=n_++1,\ldots,n_++n_-$. The spaces $\mathbf
D^{n_-}_{n_+}$ possess a natural grading which is determined by that of the
Grassmann algebra. The parity of an element $f$ of these spaces is denoted by
$\varepsilon(f)$. We also set $\varepsilon_A=0$ for $A=1,\ldots, n_+$ and
$\varepsilon_A=1$ for $A=n_++1,\ldots, n_++n_-$.

The integral on $\mathbf D^{n_-}_{n_+}$ is defined by the relation $ \int \di
z\, f(z)= \int_{\R^{n_+}}\di x\int \di\xi\, f(z), $ where the integral on the
Grassmann algebra is normed by the condition $\int \di\xi\,
\xi^1\ldots\xi^{n_-}=1$. We identify $\G^{n_-}$ with its dual space
$\G^{\prime n_-}$ setting $f(g)=\int\di\xi\, f(\xi)g(\xi)$, $f,g\in
\G^{n_-}$.  Correspondingly, the space $\mathbf D^{\prime n_-}_{n_+}$ of
continuous linear functionals on $\mathbf D^{n_-}_{n_+}$ is identified  with
the space $\D^\prime(\R^{n_+})\otimes \G^{n_-}$.  As a rule, the value $m(f)$
of a functional $m\in \mathbf D^{\prime n_-}_{n_+}$ on a test function $f\in
\mathbf D^{n_-}_{n_+}$ will be written in the ``integral'' form:
$m(f)=\int \di z\, m(z) f(z)$.

Let $L$ be a superalgebra of functions $f(z)\in\mathbf D^{n_-}_{n_+}$ with
the usual ``pointwise'' product.

Consider an algebra ${\cal A}=\sum\limits_\oplus{\cal A}_k$, where
${\cal A}_k$ is the vector space of even $k$-linear separately continuous
forms $\Phi_k(z|f_1,\ldots,f_k)\in{\cal A}_k$, $k=0,1,2,\ldots$
taking value in $\mathbf D^{n_-}_{n_+}$,
$\varepsilon(\Phi_k(z|f_1,\ldots,f_k))=\sum_{i=1}^k\varepsilon(f_i)$,
${\cal A}_0 \subset \mathbf D^{n_-}_{n_+}$,
$\Phi_0(z)=\phi(z)\in\mathbf D^{n_-}_{n_+}$,
$\varepsilon_{\Phi_0}=\varepsilon(\phi)=0$,
with (noncommutative) associative product
$$
\left( \Phi_k\diamond\Psi_p\right)
(z|f_1,\ldots,f_{k+p})=\Phi_k(z|f_1,\ldots,f_k)
\Psi_p(z|f_{k+1},\ldots,f_{k+p}).
$$
This algebra has a natural grading $g$: $g(\Phi_k)=k$, $g({\cal A}_k)=k$, and
the Hohshild differential $d_H$: ${\cal A}_k\rightarrow{\cal A}_{k+1}$,
$g(d_H)=1$, which acts by the rule:
\bee
d_{k}\Phi_k(z|f_1,\ldots,f_{k+1})&=&f_1(z)\Phi_k(z|f_2,\ldots,f_{k+1})
\nn
&+&\sum\limits_{i=1}^k(-1)^i
\Phi_k(z|f_1,\ldots,f_{i-1},f_if_{i+1},f_{i+2},\ldots,f_{k+1})
\nn
&+&
(-1)^{k+1}\Phi_k(z|f_1,\ldots,f_k)f_{k+1}(z),
\eee
Here $d_k \stackrel {def} = d_H|_{{\cal A}_k}$. Differential $d_H$ has the following
evident properties
\bee
&&d_{0}\Phi_0(z|f)=f(z)\phi(z)-\phi(z)f(z)=0,
\nn
&&d_{k+p}(\Phi_k\Psi_p)=(d_{k}\Phi_k)\Psi_p+(-1)^{g(\Phi_k)}\Phi_kd_{p}\Psi_p,
\nn
&&d_{k+1}d_{k}=0,\;k=0,1,....
\eee

Zeroth, first and second Hochschild cohomologies are found in successive sections.

\section{$H^0$}
The cohomological equation
\[
d_{0}\Phi_0(z|f)=0
\]
is satisfied identically for any forms $\Phi_0(z)$, such that we have
$H^0={\cal A}_0$.

\section{$H^1$}
In this case the cohomological equation has the form
\begin{equation}
f(z)\Phi_1(z|g)-\Phi_1(z|fg)+\Phi_1(z|f)g(z)=0. \label{H.3.1}
\end{equation}
Let%
\footnote
{Let $z=(x,\xi)$. Here and below we use the notation $z\cup U$ ($z\cap U$) instead of
$V_z\cup U$ ($V_z\cap U$) where $V_z$ is some neighborhood
of $x$.}
\be\label{dom1}
[z\cup\mathrm{supp}(g)]\cap\mathrm{supp}(f)=\varnothing.
\ee
We obtain from Eq. (\ref{H.3.1})
\[
\hat{\Phi}_1(z|f)=0,
\]
from what it follows%
\footnote
{details can be found in \cite{SKT1}.}
\[
\Phi_1(z|f)=\sum\limits_{q=0}^Qm(z)^{(A)_q}(\partial_A)^qf(z).
\]
Here $\hat{\Phi}_1(z|f)$ means the restriction of the linear form
${\Phi}_1(z|f)$ on the domain
(\ref{dom1}) or, equivalently, on the domain
$z\cap \mathrm{supp}(f) =\varnothing$,

Analogously to method used in \cite{Zh}, \cite{SKT1} let us
choose $f(z)=e^{zp}$, $g(z)=e^{zk}$ in some neighborhood of $x$.
We obtain from Eq. (\ref{H.3.1})
\begin{equation}
F(z|k)-F(z|p+k)+F(z|p)=0, \label{H.3.2}
\end{equation}
where
\[
F(z|p)=\sum\limits_{q=0}^Qm(z)^{(A)_q}(p_A)^q.
\]
Eq. (\ref{H.3.2}) gives
\[
F(z|p)=m(z)^Ap_A
\]
or
\[
\Phi_1(z|f)=m(z)^A\partial_Af(z).
\]
It is obvious that the cohomologies with different $m(z)^A$ are independent.

\section{$H^2$}
The cohomological equation has the form
\begin{equation}\label{H.4.1}
f(z)\Phi_2(z|g,h)-\Phi_2(z|fg,h)+\Phi_2(z|f,gh)-\Phi_2(z|f,g)h(z)=0.
\end{equation}

Let ${\cal L}_k \subset {\cal A}_k$  be
the spaces of local
forms. The local form is such
$\Phi_k \in {\cal A}_k$ that
if
$z\cap\mathrm{supp}(f_i)=\varnothing$ for some $1\leq i\leq k$
then
$\Phi_k(z|f_1,...,f_k)=0$.

\subsection{Nonlocal part of cocycle}
Let
\be\label{dom2}
[z\cup\mathrm{supp}(h)]\cap\mathrm{supp}(f)=
[z\cup\mathrm{supp}(h)]\cap\mathrm{supp}(g)=
\mathrm{supp}(f)\cap\mathrm{supp}(g)=\varnothing.
\ee
We obtain from Eq. (\ref{H.4.1})
\[
\hat{\Phi}_2(z|f,g)=0,
\]
which yields
\begin{equation}\label{H.4.1.1}
\Phi_2(z|f,g)=\sum\limits_{q=0}^Q\{m^{(A)_q}_1(z|f)(\partial_A)^qg(z)+
[(\partial_A)^qf(z)]m^{(A)_q}_2(z|g)+m^{(A)_q}_3(z|f(\partial_A)^qg)\}.
\end{equation}
Here notation $\hat{\Phi}_2(z|f,g)$ is used for nondiagonal part of
${\Phi}_2(z|f,g)$, i.e. for the restriction of $\Phi$ on the domain (\ref{dom2})
or on the domain
$z\cap \mathrm{supp}(f)=z\cap \mathrm{supp}(g)=
\mathrm{supp}(f)\cap\mathrm{supp}(g)=\varnothing$.

Let us note that all the forms $m^{(A)_q}_a$, and the form $m^{(A)_0}_1$
particularly, are
globally defined distributions.

Consider the domain
\be\label{dom3}
[z\cup\mathrm{supp}(f)\cup\mathrm{supp}(h)]\cap\mathrm{supp}(g)=
\varnothing.
\ee
We obtain from Eq. (\ref{H.4.1}) the equation for the restriction of
the form $\Phi$ on the domain (\ref{dom3})
\begin{equation}\label{H.4.1.2}
f(z)\hat{\Phi}_2(z|g,h)-\hat{\Phi}_2(z|f,g)h(z)=0.
\end{equation}
Substituting representation (\ref{H.4.1.1}) in Eq. (\ref{H.4.1.2}), we find
\[
\sum\limits_{q=0}^Q\{f(z)\hat{m}^{(A)_q}_1(z|g)(\partial_A)^qh(z)-
[(\partial_A)^qf(z)]\hat{m}^{(A)_q}_2(z|g)h(z)\}=0,
\]
from what it follows
\begin{eqnarray*}
&&m^{(A)_q}_1(z|g)\in {\cal L}_1,
\quad m^{(A)_q}_2(z|g)\in {\cal L}_1,
\quad q\geq1, \\
&&m^{(A)_0}_2(z|g)-m^{(A)_0}_1(z|g)\in {\cal L}_1.
\end{eqnarray*}

So, the form $\Phi_2(z|f,g)$ can be represented in the form
\begin{eqnarray*}
&&\Phi_2(z|f,g)=[\sum\limits_{q=0}^Qm^{(A)_q}_3(z|f(\partial_A)^qg)+
m^{(A)_0}_1(z|fg)]+ \\
&&+[f(z)m^{(A)_0}_1(z|g)-m^{(A)_0}_1(z|fg)+m^{(A)_0}_1(z|f)g(z)]+
\Phi_\mathrm{loc}(z|f,g), \\
&&\Phi_\mathrm{loc}(z|f,g)\in {\cal L}_2
\end{eqnarray*}
or, redefining $m^{(A)_0}_3(z|f)$,
\begin{eqnarray}
\!\!\!\!&&\Phi_2(z|f,g)\!=\Phi_{2|3}(z|f,g)\!+d_{1}\Phi_{1|1}(z|f,g)\!+
\Phi_\mathrm{loc}(z|f,g),\;\Phi_{1|1}(z|f)=m^{(A)_0}_1\!(z|f),\;
\varepsilon_{\Phi_{1|1}}\!=0, \label{H.4.1.3} \\
\!\!\!\!&&\Phi_{2|3}(z|f,g)=\sum\limits_{q=0}^Qm^{(A)_q}_3(z|f(\partial_A)^qg).
\label{H.4.1.4}
\end{eqnarray}

To specify $\Phi_{2|3}$, consider the domain
\bee\label{dom4}
z\cap[\mathrm{supp}(f)\cup\mathrm{supp}(g)\cup\mathrm{supp}(h)]=
\varnothing.
\eee
Using representation (\ref{H.4.1.3}), we obtain from Eq. (\ref{H.4.1})
\begin{equation}\label{H.4.1.5}
\hat{\Phi}_{2|3}(z|fg,h)-\hat{\Phi}_{2|3}(z|f,gh)=0.
\end{equation}
Substituting representation (\ref{H.4.1.4}) in Eq. (\ref{H.4.1.5}), we find
\[
\sum\limits_{q=0}^Q\hat{m}^{(A)_q}_3(z|\{fg(\partial_A)^qh-
f(\partial_A)^q(gh)\})=0,
\]
Choosing $g(z)=e^{-zp}$, $h(z)=e^{zp}$ on $\mathrm{supp}(f)$, we obtain
\[
\hat{F}(z|f;p)-\hat{F}(z|f;0)=0,\quad
F(z|f;p)=\sum\limits_{q=0}^Q\hat{m}^{(A)_q}_3(z|f)(p_A)^q,
\]
and then
\[
\hat{m}^{(A)_q}_3(z|f)=0\;\Rightarrow\; m^{(A)_q}_3(z|f)\in {\cal L}_1,
\quad q\geq1.
\]

So, the form $\Phi_2(z|f,g)$ can be represented in the form
\begin{equation}
\Phi_2(z|f,g)=m(z|fg)+d_{1}\Phi_{1|1}(z|f,g)+\Phi_\mathrm{loc}(z|f,g),\quad
m(z|f)=m^{(A)_0}_3(z|f). \label{H.4.1.6}
\end{equation}

At last, consider the domain
\bee\label{dom5}
[z\cup\mathrm{supp}(h)]\cap[\mathrm{supp}(f)\cup\mathrm{supp}(g)]=
\varnothing.
\eee
Using representation (\ref{H.4.1.6}), we obtain from Eq. (\ref{H.4.1})
\[
m(z|fgh)+m(z|fg)h(z)=0
 \]
and
\[
m(z|f)=0.
\]

Finally, we have obtained
\begin{eqnarray}\label{H.4.1.7}
&&\Phi_2(z|f,g)=\Phi_{2|\mathrm{loc}}(z|f,g)+d_{1}\Phi_{1|1}(z|f,g), \\
&&\Phi_{2|\mathrm{loc}}(z|f,g)=\sum\limits_{k,l=0}^Q
f(z)(\overleftarrow{\partial_A})^km^{(A)_k|(B)_l}(z)
(\partial_B)^lg(z),\quad
\varepsilon_{m^{(A)_k|(B)_l}}=\varepsilon_{A_1}+\cdots+\varepsilon_{B_l},
\nonumber \\
&&(\partial_A)^k=\partial_{A_k}\cdots\partial_{A_1}, \;
(\overleftarrow{p_A})^k=p_{A_k}\cdots p_{A_1},\;\;
(\overleftarrow{\partial_A})^k=
\overleftarrow{\partial_{A_1}}\cdots\overleftarrow{\partial_{A_k}},\;
(p_A)^k=p_{A_1}\cdots p_{A_k}, \nonumber \\
&&m^{(A)_k}=m^{A_1\cdots A_k},\;\;m^{\cdots A_iA_{i+1}\cdots}=
(-1)^{\varepsilon_{A_i}\varepsilon_{A_{i+1}}}m^{\cdots A_{i+1}A_i\cdots},
\nonumber
\end{eqnarray}
where all the coefficients $m^{(A)_k|(B)_l}(z)$
and the form $\Phi_{1|1}(z|f)$ are defined globally.

One can see that nonlocal parts of cocycles are exact forms.

\subsection{Local cocycles}
For local form, the cohomological equation (\ref{H.4.1}) takes the form
\begin{eqnarray}
\!\!\!\!\!\!\!\!\!\!\!&&\sum\limits_{k,l=0}^Q
\left(f(z)[g(z)(\overleftarrow{\partial_A})^k]
m^{(A)_k|(B)_l}(z)(\partial_B)^lh(z)-[f(z)g(x)](\overleftarrow{\partial_A})^k
m^{(A)_k|(B)_l}(z)(\partial_B)^lh(z)+\right. \nonumber \\
\!\!\!\!\!\!\!\!\!\!\!&&+\!\left.f(z)(\overleftarrow{\partial_A})^k
m^{(A)_k|(B)_l}(z)(\partial_B)^l[g(z)h(z)]
\!-\!f(z)(\overleftarrow{\partial_A})^k
m^{(A)_k|(B)_l}(z)[(\partial_B)^lg(z)]h(z)=0\!\right)\!\!. \label{H.4.2.1}
\end{eqnarray}
Let $f(z)=e^{pz}$, $g(z)=e^{qz}$, $h(z)=e^{rz}$ in some neighborhood of $x$.
Then the cohomological equation
transforms to the form
\begin{eqnarray}\label{H.4.2.2}
&&F(z|q,\tilde{r})-F(z|p+q,\tilde{r})+F(z|p,\tilde{q}+\tilde{r})-
F(z|p,\tilde{q})=0, \\
&&F(z|p,q)=\sum\limits_{k,l=0}^Q(p_A)^km^{(A)_k|(B)_l}(z)(q_B)^l, \;\;
\tilde{p}_A=(-1)^{\varepsilon_A}p_A. \nonumber
\end{eqnarray}

Let us apply the operator
$\left.\partial/\partial p_A\right|_{p=0}$ to Eq. (\ref{H.4.2.2}). We obtain
\begin{eqnarray}
&&\frac{\partial}{\partial q_A}F(z|q,\tilde{r})=\Psi^A(z|q+r)-\Psi^A(z|q),
\label{H.4.2.3} \\
&&\Psi^A(z|q)=\left.\frac{\partial}{\partial p_A}F(z|p,\tilde{q})\right|_{p=0}.
\nonumber
\end{eqnarray}
It follows from Eq. (\ref{H.4.2.3})
\begin{equation}\label{H.4.2.4}
\frac{\partial}{\partial q_A}\Psi^B(z|q+r)-
(-1)^{\varepsilon_A\varepsilon_B}
\frac{\partial}{\partial q_B}\Psi^A(z|q+r)=
\frac{\partial}{\partial q_A}\Psi^B(z|q)-
(-1)^{\varepsilon_A\varepsilon_B}
\frac{\partial}{\partial q_B}\Psi^A(z|q)
\end{equation}
and then
\begin{eqnarray}\label{H.4.2.5}
&&\frac{\partial}{\partial r_A}\Psi^B(z|r)-
(-1)^{\varepsilon_A\varepsilon_B}
\frac{\partial}{\partial r_B}\Psi^A(z|r)=-
(-1)^{\varepsilon_A\varepsilon_B}\omega^{AB}(z), \\
&&\omega^{AB}(z)=\left.\frac{\partial}{\partial q_B}\Psi^A(z|q)-
(-1)^{\varepsilon_A\varepsilon_B}
\frac{\partial}{\partial q_A}\Psi^B(z|q)\right|_{q=0}=
-(-1)^{\varepsilon_A\varepsilon_B}\omega^{BA}(z). \nonumber
\end{eqnarray}

A general solution of Eq. (\ref{H.4.2.5}) is
\begin{equation}\label{H.4.2.6}
\Psi^A(z|r)=\frac{1}{2}\omega^{AB}(z)r_B+
\frac{\partial}{\partial r_A}\phi(z|r).
\end{equation}
Function (\ref{H.4.2.6}) satisfies Eq. (\ref{H.4.2.4}) also. Substituting
Exp. (\ref{H.4.2.6}) in Eq. (\ref{H.4.2.3}), we obtain
\[
\frac{\partial}{\partial q_A}\left(F(z|q,\tilde{r})-
\frac{1}{2}q_A\omega^{AB}(z)r_B-\phi(z|q+r)+\phi(z|q)\right)=0
\]
and as a result
\begin{equation}\label{H.4.2.7}
F(z|q,r)=\frac{1}{2}q_A\omega^{AB}(z)\tilde{r}_B+\phi(z|q+\tilde{r})
-\phi(z|q)+\varphi(z|\tilde{r}).
\end{equation}
Substituting Exp. (\ref{H.4.2.7}) in Eq. (\ref{H.4.2.2}) we find
that function (\ref{H.4.2.7}) satisfies Eq. (\ref{H.4.2.2}) if function
$\varphi(z|r)$ is equal to $\varphi(z|r)=-\phi(z|r)-\varphi_1(z)$. So, we
obtain
\[
F(z|p,q)=\frac{1}{2}p_A\omega^{AB}(z)\tilde{q}_B+\phi(z|p+\tilde{q})
-\phi(z|p)-\phi(z|\tilde{q}),
\]
where a redefinition $\phi(z|p)\rightarrow\phi(z|p)+\varphi_1(z)$ was made, or
\begin{eqnarray*}
&&\Phi_{2\mathrm{loc}}(z|f,g)=\frac{1}{2}f(z)\overleftarrow{\partial_A}
\omega^{AB}(z)\partial_Bg(z)+d_{1}\Phi_{1|2}(z|f,g), \\
&&\Phi_{1|2}(z|f)=-f(z)\sum\limits_{k=0}^K
(\overleftarrow{\partial_A})^k\phi^{(A)_k}(z),
\end{eqnarray*}
where $\phi^{(A)_k}(z)$ are coefficients of the polynomial $\phi(z|p)$,
$\phi(z|p)=\sum\limits_{k=0}^K(p_A)^k\phi^{(A)_k}(z)$.

Finally, with (\ref{H.4.1.7}) taken into account, we find that general
solution of cohomological equation (\ref{H.4.1}) has the form
\begin{eqnarray}\label{H.4.2.7a}
&&\Phi_2(z|f,g)=m_\omega(z|f,g)+d_{1}\Phi_1(z|f,g), \\
&&m_\omega(z|f,g)=\frac{1}{2}f(z)\overleftarrow{\partial_A}
\omega^{AB}(z)\partial_Bg(z)=-
(-1)^{\varepsilon(f)\varepsilon(g)}m_\omega(z|g,f),\;\;
\varepsilon(\omega^{AB})=\varepsilon_A+\varepsilon_B, \nonumber \\
&&\Phi_1(z|f)=\Phi_{1|1}(z|f,g)+\Phi_{1|2}(z|f). \nonumber
\end{eqnarray}

The cohomologies with different $\omega^{AB}(z)$ are independent. Indeed,
consider an equation
\begin{equation}\label{H.4.2.8}
f(z)\overleftarrow{\partial_A}\omega^{AB}(z)\partial_Bg(z)=
d_{1}\Psi_1(z|f,g)=f(z)\Psi_1(z|g)-\Psi_1(z|fg)+\Psi_1(z|f)g(z).
\end{equation}
Let
\[
[z\cup\mathrm{supp}(g)]\cap\mathrm{supp}(f)=\varnothing.
\]
We find from Eq. (\ref{H.4.2.8})
\[
\hat{\Psi}_1(z|f)=0,
\]
i.e. the form $\Psi_1(z|f)$ is local:
\[
\Psi_1(z|f)=f(z)\sum\limits_{k=0}^K(\overleftarrow{\partial_A})^k
\psi^{(A)_k}(z)\equiv f(z)\psi(z|\overleftarrow{\partial}).
\]
Let $f(z)=e^{pz}$, $g(z)=e^{qz}$. It follows from Eq. (\ref{H.4.2.8})
\[
p_A\omega^{AB}(z)(-1)^{\varepsilon_B}q_B=\psi(z|q)-\psi(z|p+q)+\psi(z|p).
\]
R.h.s. of this equation is symmetric under exchange $p\leftrightarrow q$,
such that we have
\[
p_A\omega^{AB}(z)(-1)^{\varepsilon_B}q_B=
q_B\omega^{BA}(z)(-1)^{\varepsilon_A}p_A=-
p_A\omega^{AB}(z)(-1)^{\varepsilon_B}q_B
\]
from what it follows $\omega^{AB}(z)=0$, that is, Eq. (\ref{H.4.2.8}) has
solutions only for $\omega^{AB}(z)=0$.


\begin{thebibliography}{99}

\bibitem{1}
{\it F.~Bayen, M.~Flato, C.~Fronsdal, A.~Lichnerovich and
D.~Sternheimer}, Ann.Phys., {\bf 111}, 61 (1978);
Ann.Phys., {\bf 111}, 111 (1978).

\bibitem{KarMas}
{\it M.~V.~Karasev and V.~P.~Maslov}, Nonlinear Poisson brackets. Geometry
and Quantization [in Russian], Nauka, Moscow (1991);
{\it M.~V.~Karasev and V.~P.~Maslov},
Nonlinear Poisson brackets. Geometry and
Qantization, AMS, Providence, RI (1993).

\bibitem{Fed3}
{\it B.~Fedosov}, Deformation quantization and Index Theory, Akademie Verlag,
Berlin, 1996.

\bibitem{Kon}
{\it M.~Kontsevich}, Deformation quantization of Poisson manifolds, I,
q--alg/9709040.

\bibitem{Nek}
{\it N.~Nekrasov}, Trieste Lectures on solitons in noncommutative gauge
theories, hep--th/0011095.

\bibitem{AS}
{\it A.~Schwarz}, Gauge theories on noncommutative spaces, hep--th/0011261.

\bibitem{Gro}
{\it H.~Groenewold}, Physica, {\bf 12}, 405 (1946).

\bibitem{Ber}
{\it F.~A.~Berezin}, Uspekhi Fiz. Nauk, {\bf 23}  (1980).

\bibitem{Tu}
{\it I.~V.~Tyutin},
The general form of the star--product on the Grassmann algebra,
hep-th/0101046.

\bibitem{Leites}
{\it D.~A.~Leites and I.~M.~Shchepochkina},
How to quantize the antibracket,
Theor.~Math.~Phys., {\bf 126}, 281 (2001).

\bibitem{TK}
{\it S.~E.~Konstein, I.~V.~Tyutin},
General form of the deformation
of Poisson superbracket
on (2,2)-dimensional superspace, hep-th/061206.

\bibitem{Zh1}
{\it V.~V.~Zharinov},
Hochschild cohomologies of the algebra of smooth functions,
Theor.~Math.~Phys., {\bf 140}, 1195 (2004).

\bibitem{SKT1}
{\it S.~E.~Konstein, A.~G.~Smirnov and I.~V.~Tyutin,}
Cohomologies of the Poisson superalgebra,
Theor.~Math.~Phys., {\bf 143}, 625 (2005);
hep-th/0312109.

\bibitem{Zh}
{\it V.~V.~Zharinov,}  Theor.~Math.~Phys., {\bf 136}, 1049  (2003).

\end{thebibliography}
\end{document}